\documentclass[preprint,12pt]{elsarticle}

\usepackage{graphics}

\usepackage{mathrsfs}
\usepackage{amsmath}
\def\Vec#1{\mbox{\boldmath $#1$}}

\journal{Physica A}

\begin{document}

\begin{frontmatter}



\title{Irreversibility and Entropy Production \\ in Transport Phenomena II\\--Statistical-mechanical Theory on Steady States including Thermal Disturbance and Energy Supply--}


\author{Masuo Suzuki}
\address{Tokyo University of Science\\Kagurazaka 1-3, Shinjuku, Tokyo, 162-8601}
\ead{msuzuki@rs.kagu.tus.ac.jp}

\begin{abstract}
Some general aspects of nonlinear transport phenomena are discussed on the basis of two kinds of formulations obtained by extending Kubo's perturbational scheme of the density matrix and Zubarev's non-equilibrium statistical operator formulation.
Both formulations are extended up to infinite order of an external force in compact forms and their relationship is clarified through a direct transformation.
In order to make it possible to apply these formulations straight-forwardly to thermal disturbance, a mechanical formulation of it is given (in a more convenient form than Luttinger's formulation) by introducing the concept of a thermal field $\Vec{E}_{\text{T}}$ which corresponds to the temperature gradient and by defining its conjugate heat operator $\Vec{A}_{\text{H}}=\sum_{j} h_{j} \Vec{r}_j$ for a local internal energy $h_j$ of the thermal particle $j$.
This yields a transparent derivation of the thermal conductivity $\kappa $ of the Kubo form and the entropy production $(dS/dt)_{\text{irr}}=\kappa E^2_{\text{T}}/T$.
Mathematical aspects on the non-equilibrium density-matrix will also be discussed.
In Paper I (Physica A {\bf{390}}(2011)1904), the symmetry-separated von Neumann equation with relaxation terms extracting generated heat outside the system was introduced to describe the steady state  of the system.
In this formulation of the steady state, the internal energy $\left<\mathcal{H}_0\right>_t$ is time-independent but the field energy $\left<\mathcal{H}_1\right>_t(=-\left<\Vec{A}\right>_t\cdot \Vec{F})$ decreases as time $t$ increases.
To overcome this problem, such a statistical mechanical formulation is proposed here as includes energy supply to the system from outside by extending the symmetry-separated von Neumann equation given in Paper I.
This yields a general theory based on the density-matrix formulation on a steady state with energy supply inside  and heat extraction outside and consequently with both $\left< \mathcal{H}_0 \right>_t$ and $\left< \mathcal{H}_1 \right>_t$ constant.
Furthermore, this steady state gives a positive entropy production. 

The present general formulation of the current yields a compact expression of the time derivative of entropy production, which yields the plausible justification of the principle of minimum entropy production in the steady state even for nonlinear responses.
\end{abstract}


\end{frontmatter}

\section{Introduction}
In the first paper of this series \cite{1}, the present author has derived the irreversible entropy production in transport phenomena [1-16] starting from the von Neumann equation, as Kubo et al. [2-4] formulated the linear response scheme.
A new aspect of the present author's theory \cite{1} is that the entropy operator $\mathcal{S}$ is defined using the equilibrium density matrix $\rho_{\text{eq}}$ as
\begin{equation}
\mathcal{S}=-k_{\text{B}}\log \rho_{\text{eq}}=(\mathcal{H}_0-\mathcal{F}_0)/T\label{eq:1}
\end{equation}
with $\mathcal{F}_0=-k_{\text{B}}T\log \text{Tr}\exp(-\beta \mathcal{H}_0)$ and that the entropy production is obtained thereby from the symmetric part $\rho_{\text{s}}(t)$ of the density matrix\cite{1}:
\begin{equation}
\left(\frac{dS}{dt}\right)_{\text{irr}}=\frac{d}{dt}\mathcal{S}\rho(t)=\frac{1}{T}\frac{d}{dt}\text{Tr}\mathcal{H}_0\rho_{\text{s}}(t)=\frac{\sigma_{F} F^2}{T};\quad F=|\Vec{F}|,\label{eq:2}
\end{equation}
Here, $\mathcal{H}_0$ denotes the Hamiltonian of internal energy,
$\Vec{F}$ a static field, $\sigma_{F}$ the nonlinear transport coefficient and $T$ the temperature.
Thus, the dissipation (or Joule heat in electric conduction) $\sigma_{F} F^2$ is obtained\cite{1} from the irreversible entropy production as $\sigma_{F}F^2=T(d\mathcal{S}/dt)_{\text{irr}}$ even in the nonlinear regime as well as in the linear scheme.
This is a big contrast to the traditional phenomenological formulation of dissipation using the complex admittance.
On the other hand, the current $\Vec{J}_F$ is given by $\Vec{J}_F=\text{Tr}\Vec{j}\rho_{\text{a}}(t)=\sigma_F\Vec{F}$ using the antisymmetric part $\rho_{\text{a}}(t)$ and in particular the Kubo formula is given by the first-order term $\rho_1(t)$ of the antisymmetric part $\rho_{\text{a}}(t)$ as
\begin{equation}
\Vec{J}=\text{Tr}\Vec{j}\rho_1(t)=\sigma_0\Vec{F};\quad
\sigma_0=\int_0^{\infty}dt\int_0^{\beta}d\lambda e^{-\epsilon t}\left< \Vec{j}\Vec{j}(t+i\hbar \lambda ) \right>_0\label{eq:3}
\end{equation}
with the adiabatic factor $e^{-\epsilon t}$.
Here, the density matrix $\rho(t)$ is separated more explicitly into the two parts $\rho_{\text{s}}(t)$ and $\rho_{\text{a}}(t)$ defined by 
\begin{subequations}
\begin{equation}
\rho_{\text{s}}(t)=\rho_0+\rho_2(t)+\rho_4(t)+\cdots, \label{eq:4a}
\end{equation}
and
\begin{equation}
\rho_{\text{a}}(t)=\rho_1(t)+\rho_3(t)+\rho_5(t)+\cdots, \label{eq:4b}
\end{equation}
\end{subequations}
as the power-series expansion with respect to the field $\Vec{F}(t)$ defined in the Hamiltonian
\begin{equation}
\mathcal{H}(t)=\mathcal{H}_0+\mathcal{H}_1(t);\quad \mathcal{H}_1(t)=-\Vec{A}\cdot\Vec{F}(t).\label{eq:5}
\end{equation}

One of the purposes of the present paper is to apply the general theory \cite{1} to the thermal conduction by constructing a mechanical formulation of the thermal conduction.
For this purpose, the concept of a thermal field $\Vec{E}_{\text{T}}$ and its conjugate heat operator $\Vec{A}_{\text{H}}$ will be introduced in Section 5.
This formulation yields a new direct derivation of the heat current operator $\Vec{j}_{\text{H}}$ in the Kubo formula (\ref{eq:3}).

Before presenting the mechanical formulation of thermal conductivity, it will be instructive to formulate nonlinear responses in two ways, namely using the Kubo-type scheme and by extending the Zubarev-type scheme up to infinite order of the field $\Vec{F}(t)$.
The relationship between these schemes will be studied.
The unboundness of the relevant non-equilibrium operators obtained in the present paper will be also discussed.

Finally the idea of the symmetry separation in the von Neumann equation is shown to be useful in formulating the energy supply to the relevant system, without changing the previous formulation\cite{1} of the internal energy $\left< \mathcal{H}_0 \right>_t$, current $\Vec{J}_F$ and irreversible entropy production $(dS/dt)_{\text{irr}}$.
For this purpose, we introduce a new concept of the stationary density-matrix whose time derivative can be allowed to be not only zero but also a non-vanishing constant operator.

\section{Kubo-type formulation of the non-equilibrium density matrix}
In the previous paper \cite{1}, the author studied the irreversibility and entropy production in transport phenomena [2-16], using the idea of the symmetry separation of the density matrix in the von Neumann equation.

It will be instructive to discuss explicitly the nonlinear solution of the von Neumann equation
\begin{equation}
\frac{\partial }{\partial t}\rho(t)=\frac{1}{i\hbar}[\mathcal{H}(t),\rho(t)] ;\quad \mathcal{H}(t)=\mathcal{H}_0+\mathcal{H}_1(t)\text{ and }\mathcal{H}_1(t)=-\Vec{A}\cdot\Vec{F}(t)\label{eq:6}
\end{equation}
for a field $\Vec{F}(t)$.
Following Kubo\cite{2}, we may put $\rho(t)=\rho_0+\Delta\rho(t)$ with 
\begin{equation}
\rho_0=e^{-\beta \mathcal{H}_0}/Z_0(\beta );\quad Z_0(\beta )=\text{Tr}e^{-\beta \mathcal{H}_0}.\label{eq:7}
\end{equation}
The solution $\Delta \rho(t)$ satisfying the equation 
\begin{equation}
\frac{\partial}{\partial t}\Delta \rho(t)=\frac{1}{i\hbar}[\mathcal{H}(t),\Delta \rho(t)]+\frac{1}{i\hbar}[\mathcal{H}_1(t),\rho_0 ]\label{eq:8}
\end{equation}
is given formally by
\begin{equation}
\Delta \rho(t)=\frac{1}{i\hbar}\int_{t_0}^{t}\mathcal{U}(t,t')[\mathcal{H}_1(t'),\rho_0]\mathcal{U}^{-1}(t,t')dt',\label{eq:9}
\end{equation}
where $\mathcal{U}(t,t')$ is the ordered exponential defined by [17-21]
\begin{align}
\mathcal{U}(t,t')&=\exp_{+}\left(\frac{1}{i\hbar} \int _{t'}^{t}\mathcal{H}(s)ds\right)\notag
\\
&=1+\frac{1}{i\hbar}\int _{t'}^{t}\mathcal{H}(t_1)dt_1+\left(\frac{1}{i\hbar}\right)^2\int_{t'}^{t}dt_1\int_{t'}^{t_1}dt_2\mathcal{H}(t_1)\mathcal{H}(t_2)+\cdots\notag
\\
&+\left(\frac{1}{i\hbar}\right)^n \int_{t'}^{t}dt_1\int_{t'}^{t_1}dt_2\cdots \int_{t'}^{t_{n-1}}dt_n\mathcal{H}(t_{1})\mathcal{H}(t_{2})\cdots\mathcal{H}(t_{n})+\cdots. \label{eq:10}
\end{align}
It will be more convenient for expressing the current $\Vec{J}_F=\text{Tr}\Vec{j}\rho(t)=\text{Tr}\Vec{\dot{A}}\Delta \rho(t)$ in the Kubo-type (generalized) form \cite{22,23} to rewrite $\Delta \rho(t) $ as 
\begin{equation}
(\Delta \rho(t))_{\text{K-type}}=\int_{t_0}^{t}dt'\mathcal{U}(t,t')\left(\int_{0}^{\beta}d\lambda \rho_0\Vec{j}(-i\hbar \lambda)\right)\mathcal{U}^{-1}(t,t')\cdot\Vec{F}(t'),\label{eq:11}
\end{equation}
where
\begin{equation}
Q(t)=e^{-\frac{t}{i\hbar}\mathcal{H}_0}Qe^{\frac{t}{i\hbar}\mathcal{H}_0}\text{ and }\Vec{j}(-i\hbar\lambda)=e^{\lambda \mathcal{H}_0}\Vec{j}e^{-\lambda \mathcal{H}_0};\quad \Vec{j}=\Vec{\dot{A}}=\frac{1}{i\hbar}[\Vec{A},\mathcal{H}_0]\label{eq:12}
\end{equation}
for any operator $Q$ excluding the density matrix.
Thus, the nonlinear current $\Vec{J}_F$ is given in the form
\begin{equation}
\Vec{J}_F=\text{Tr}\Vec{j}\left(\Delta\rho(t)\right)_{\text{K-type}}=\int_{t_0}^{t}dt'\int_{0}^{\beta}d\lambda \left<\Vec{j}(-i\hbar\lambda)\mathcal{U}^{-1}(t,t')\Vec{j}\mathcal{U}(t,t')\right>_0\cdot \Vec{F}(t').\label{eq:13}
\end{equation}

If the force $\Vec{F}(t)$ contains only the adiabatic factor $e^{\epsilon t}(\epsilon\rightarrow +0)$, namely it is essentially time-independent as in Paper I, then the above expression (\ref{eq:11}) is simplified as
\begin{equation}
(\Delta \rho(t))_{\text{K-type}}=\int_{t_0}^{t}dt'\left(\int_{0}^{\beta}d\lambda \rho_0\Vec{j}(-i\hbar \lambda)\right)(t'-t;\Vec{F})\cdot\Vec{F}e^{\epsilon t'}.\label{eq:14}
\end{equation}
Here, $Q(t;\Vec{F})$ except the density matrix $\rho(t)$ is defined by 
\begin{equation}
Q(t;\Vec{F})=\exp\left(-\frac{t}{i\hbar}\mathcal{H}\right)Q\exp\left(\frac{t}{i\hbar}\mathcal{H}\right);\quad Q(t;0)=Q(t)\label{eq:15}
\end{equation}
for $\mathcal{H}=\mathcal{H}_0+\mathcal{H}_1=\mathcal{H}_0-\Vec{A}\cdot\Vec{F}$ and consequently we obtain $\Vec{J}_F=\sigma_F\Vec{F}$, where
\begin{equation}
\sigma_F=\int_0^{\infty}dte^{-\epsilon t}\int_{0}^{\beta}d\lambda\left<\Vec{j}(-i\hbar\lambda)\Vec{j}(t;\Vec{F})\right>_0\text{ };\text{ }\Vec{j}(t;\Vec{F})=e^{-\frac{t\mathcal{H}}{i\hbar}}\Vec{j}e^{\frac{t\mathcal{H}}{i\hbar}}.\label{eq:16}
\end{equation}
Clearly this is reduced to the well-known Kubo formula (\ref{eq:3}), or more generally
\begin{equation}
\Vec{J}=\text{Re}\left(\sigma_0(\omega)\Vec{F}e^{i\omega t}\right);\quad \sigma_0(\omega)=\int_{0}^{\infty}dt e^{(i\omega-\epsilon)t}\int_{0}^{\beta}d\lambda\left<\Vec{j}\Vec{j}(t+i\hbar\lambda)\right>_0\label{eq:17}
\end{equation}
in the linear response scheme\cite{2}, for the complex admittance $\sigma(\omega)$ calculated using the complex external field $\Vec{F}(t)=\Vec{F}e^{i\omega t}$ in the general expression (\ref{eq:13}).
As is well known, the first equation in Eq.(\ref{eq:17}) is shown to agree with the expression directly obtained from Eq.(\ref{eq:13}) for the hermitian Hamiltonian $\mathcal{H}_{1}(t)=-\Vec{A}\cdot\Vec{F}\cos(\omega t)$.

This scheme is a statistical-mechanical perturbation theory, namely perturbation with respect to the statistical ensemble, and it is not mechanical\cite{16} in contrast to the Zubarev-type formulation.

The above formula (\ref{eq:14}) is useful in the plausible justification of the principle of minimum entropy production in the steady state, as will be discussed in Section 9.

\section{Zubarev-type formulation}
There is another type of formulation on nonlinear responses, namely the Zubarev-type non-equilibrium density-matrix scheme\cite{6,7}.

The starting point of the present general formulation is the following expression of the non-equilibrium density matrix
\begin{align}
\rho(t)&=\mathcal{U}(t,t_0)\rho_0\mathcal{U}^{-1}(t,t_0)=\frac{1}{Z_0(\beta)}\exp\left(-\beta\mathcal{U}(t,t_0)\mathcal{H}_0\mathcal{U}^{-1}(t,t_0)\right)\notag
\\
&\equiv \frac{1}{Z_0(\beta)}\exp\left(-\beta(\mathcal{H}_0+\mathcal{R}(t,t_0)) \right).\label{eq:18}
\end{align}
The non-equilibrium operator $\mathcal{R}(t,t_0)$ defined in Eq.(\ref{eq:18}) satisfies the time-evolution equation\cite{7}
\begin{equation}
\frac{\partial}{\partial t}\mathcal{R}(t,t_0)=\frac{1}{i\hbar}[\mathcal{H}(t),\mathcal{R}(t,t_0)]+\frac{1}{i\hbar}[\mathcal{H}_1(t),\mathcal{H}_0],\label{eq:19}
\end{equation}
which is proved directly or using the quantum analysis [19-21].
It should be noted that the above equation (\ref{eq:19}) is purely mechanical, because it does not include the density matrix $\rho_0$ (or $\rho(t)$).
The source term in Eq.(\ref{eq:19}) is given in terms of the commutator $[\mathcal{H}_1(t),\mathcal{H}_0]$ instead of the commutator $[\mathcal{H}_1,\rho_0]$ in Eq.(\ref{eq:8}).
This corresponds to the Lippmann-Schwinger perturbation scheme in quantum scattering theory\cite{24}.

Now the formal solution of (\ref{eq:19}) is given by
\begin{equation}
\mathcal{R}(t,t_0)=-\int_{t_0}^{t}dt'\mathcal{U}(t,t')\Vec{j}\mathcal{U}^{-1}(t,t')\cdot\Vec{F}(t').\label{eq:20}
\end{equation}
Then, the nonlinear current $\Vec{J}_F$ is expressed by
\begin{equation}
\Vec{J}_F=\text{Tr}\Vec{j}\rho(t)=\text{Tr}\Vec{j}\left(\Delta \rho (t)\right)_{\text{Z-type}},\label{eq:21}
\end{equation}
where
\begin{align}
\left(\Delta \rho (t)\right)_{\text{Z-type}}&\equiv \rho(t)-\rho_0=-\rho_0\int_{0}^{\beta}d\lambda e^{\lambda\mathcal{H}_0}\mathcal{R}(t,t_0)e^{-\lambda (\mathcal{H}_0+\mathcal{R}(t,t_0))}\notag
\\
&=-\int_{0}^{\beta}d\lambda e^{-\lambda (\mathcal{H}_0+\mathcal{R}(t,t_0))}\mathcal{R}(t,t_0)e^{\lambda\mathcal{H}_0}\rho_0,\label{eq:22}
\end{align}
or
\begin{align}
\left(\Delta \rho (t)\right)_{\text{Z-type}}&=\frac{1}{Z_0(\beta)}\int_{0}^{1}d\mu\frac{d}{d\mu}e^{-\beta(\mathcal{H}_0+\mu\mathcal{R}(t,t_0))}\notag
\\
&=\frac{1}{Z_0(\beta)}\int_{0}^{1}d\mu\int_{0}^{\beta}d\lambda e^{-(\beta-\lambda)(\mathcal{H}_0+\mu\mathcal{R}(t,t_0))}\mathcal{R}(t,t_0)e^{-\lambda(\mathcal{H}_0+\mu\mathcal{R}(t,t_0))}\label{eq:23}
\end{align}
using the integration technique\cite{23}.
The expression (\ref{eq:22}) is convenient for the perturbation expansion with respect to the field $\Vec{F}(t)$, but it is asymmetric with respect to the $\lambda$-integral.
On the other hand, the triple integral representation (\ref{eq:23}) is symmetric with respect to the field $\Vec{F}(t)$ (namely to the $\mu$-integral).
Thus, the nonlinear current $\Vec{J}_F$ is expressed in the following two ways:
\begin{align}
\Vec{J}_F&=\int_{t_0}^{t}dt'\int_{0}^{\beta}d\lambda \left<\Vec{j}e^{-\lambda (\mathcal{H}_0+\mathcal{R}(t,t_0))}\mathcal{U}(t,t')\Vec{j}\mathcal{U}^{-1}(t,t')e^{\lambda \mathcal{H}_0} \right>_0 \Vec{F}(t')\notag
\\
&=\int_{t_0}^{t}dt'\int_{0}^{\beta}d\lambda \left< e^{\lambda \mathcal{H}_0}\mathcal{U}(t,t')\Vec{j}\mathcal{U}^{-1}(t,t')e^{-\lambda (\mathcal{H}_0+\mathcal{R}(t,t_0))}\Vec{j} \right>_0 \Vec{F}(t'),\label{eq:24}
\end{align}
or
\begin{align}
\Vec{J}_F&=\int_{t_0}^{t}dt'\int_{0}^{\beta}d\lambda \int_{0}^{1}d\mu\notag
\\
&\left< e^{-(\beta-\lambda)(\mathcal{H}_0+\mu\mathcal{R}(t,t_0))}\mathcal{U}(t,t')\Vec{j}\mathcal{U}^{-1}(t,t')e^{-\lambda(\mathcal{H}_0+\mu\mathcal{R}(t,t_0))}\Vec{j}e^{\beta\mathcal{H}_0} \right>_0\Vec{F}(t'),\label{eq:25}
\end{align}
as is easily seen from Eqs.(\ref{eq:22}) and (\ref{eq:23}), respectively.
The above expressions (\ref{eq:24}) and (\ref{eq:25}) yield immediately the transport coefficients, and particularly the new expression (\ref{eq:25}) is interesting, because the quantum effect through the $\lambda$-integral is taken into account symmetrically and gradually through the $\mu$-integral.

If the force $\Vec{F}(t)$ depends on time $t$ only through the adiabatic factor $e^{\epsilon t}(\epsilon\rightarrow +0)$, then using the nonlinear current operator $\Vec{j}(t;\Vec{F})$ defined in Eq.(\ref{eq:16}) we obtain 
\begin{equation}
\Vec{J}_F=\sigma_F\Vec{F};\quad \sigma_F=\int_{0}^{\infty}dt e^{-\epsilon t}\int_0^1 d\mu \left< \Vec{j};\Vec{j}(t;\Vec{F}) \right>_{\mu F},\label{eq:26}
\end{equation}
in the limit $t_0\rightarrow -\infty$, where
\begin{subequations}
\begin{align}
\left<\Vec{j};\Vec{j}(t;\Vec{F})\right>_{\mu F}&=\int_{0}^{\beta}d\lambda \text{Tr}e^{-(\beta-\lambda)\mathcal{H}_{(\mu)}}\Vec{j}e^{-\lambda \mathcal{H}_{(\mu)}}\Vec{j}(t;\Vec{F})/Z_0(\beta)\notag
\\
&=\int_{0}^{\beta}d\lambda \left< e^{-(\beta-\lambda)\mathcal{H}_{(\mu)}}\Vec{j}e^{-\lambda\mathcal{H}_{(\mu)}}\Vec{j}(t;\Vec{F})e^{\beta\mathcal{H}_0} \right>_0\label{eq:27a}
\end{align}
with
\begin{equation}
\mathcal{H}_{(\mu)}=\mathcal{H}_0+\mu \lim_{t_0\rightarrow -\infty}\mathcal{R}(t,t_0),\label{eq:27b}
\end{equation}
and
\begin{equation}
\mathcal{R}(t,t_0)=-\int_{t_0}^{t}dt'e^{\epsilon t'}e^{\frac{(t-t')}{i\hbar}\mathcal{H}}\Vec{j}e^{-\frac{(t-t')}{i\hbar}\mathcal{H}}\cdot\Vec{F}.\label{eq:27c}
\end{equation}
\end{subequations}
By the way, the Zubarev-type density-matrix is given in the form
\begin{subequations}
\begin{equation}
\rho_{\text{Z-type}}=\rho(0) =\exp\left(-\beta (\mathcal{H}_0+\mathcal{R}_F)\right)/Z_0(\beta),\label{eq:28a}
\end{equation}
using $\mathcal{R}_F=\mathcal{R}_F(0,-\infty)=\mathcal{R}_F(t,-\infty)$, which shows  $\mathcal{U}(t,0)(\mathcal{H}_0+\mathcal{R}_F)\mathcal{U}^{-1}(t,0)=\mathcal{H}_0+\mathcal{R}_F$ and consequently we have $[\mathcal{H}_0+\mathcal{R}_F,\mathcal{H}]=0$.
(Note that this equation can not determine $\mathcal{R}_F$ uniquely.)
Here, from Eq.(\ref{eq:27c}) with Eq.(\ref{eq:16}), $\mathcal{R}_F$ is expressed by the integral
\begin{equation}
\mathcal{R}_F=-\int_0^{\infty}e^{-\epsilon t}\Vec{j}(-t;\Vec{F})\cdot\Vec{F}dt.\label{eq:28b}
\end{equation}
\end{subequations}
Now we make here a linear approximation of $\mathcal{R}_F$ with respect to the force $\Vec{F}$ to obtain
\begin{equation}
\mathcal{R}_1=-\int_{0}^{\infty}e^{-\epsilon t}\Vec{j}(-t)\cdot\Vec{F}dt\quad \text{with}\quad \Vec{j}(t)=e^{-\frac{t}{i\hbar}\mathcal{H}_0}\Vec{j}e^{\frac{t}{i\hbar}\mathcal{H}_0},\label{eq:29}
\end{equation}
if it exists in the limit $t_0\rightarrow -\infty$.
Then we obtain Zubarev's non-equilibrium statistical operator of the form
\begin{align}
\rho_{\text{Zub}}&=\exp \left( -\beta(\mathcal{H}_0+\mathcal{R}_1) \right)/Z_0(\beta)\notag
\\
&=\exp\left(-\beta \mathcal{H}_0 +\beta\int_{0}^{\infty}e^{-\epsilon t}\Vec{j}(-t)\cdot \Vec{F}dt \right)/Z_0(\beta).\label{eq:30}
\end{align}
Note that $[\mathcal{H}_0+\mathcal{R}_1,\mathcal{H}]=O(F^2)$.
If we expand Eq.(\ref{eq:30}) again up to the linear order, then we obtain the Kubo formula (\ref{eq:3}).
The above density matrices $(\Delta \rho )_{\text{K-type}} $ and $(\Delta \rho )_{\text{Z-type}} $ (or $\rho _{\text{Z-type}} $ in Eq.(\ref{eq:28a})) are useful in studying a current operator, but we have to be careful in discussing the time change of physical operators such as $\mathcal{H}_0, \mathcal{H}_1$ and entropy, in which the formula
\begin{equation}
\frac{d}{dt}\left< Q \right>_t =\frac{1}{i\hbar}\text{Tr} [Q,\mathcal{H}]\rho(0)\label{eq:31}
\end{equation}
should be used in the limit $t_0\rightarrow -\infty$ for any physical quantity $Q$, when the system is described by the original von Neumann equation (\ref{eq:6}) with $\mathcal{H}(t)=\mathcal{H}$.
The relation between the above Zubarev-type expression (\ref{eq:23}) and the Kubo-type expression (\ref{eq:11}) will be discussed in the succeeding section.

\section{Relationship between the Kubo-type and Zubarev-type formulations}
As has been shown in Sections 2 and 3, the Kubo-type expression (\ref{eq:11}) is of convolution-type, and the Zubarev-type expression (\ref{eq:23}) looks very different from Eq.(\ref{eq:11}).
It will be instructive to prove the equivalence of the two expressions by a direct transformation.
In order to realize clearly the procedure of the direct transformation, we explain the case of the time-independent non-equilibrium Hamiltonian $\mathcal{H}$ as follows.
First note that
\begin{align}
\left(\Delta \rho(t)\right)_{\text{K-type}}
&=\int_{t_0}^{t}dt'\left(\exp\left(\frac{(t-t')}{i\hbar}\delta_{\mathcal{H}}\right)\right)\frac{1}{i\hbar}\delta_{\mathcal{H}_1}\rho_0\notag
\\
&=\left(\exp\left(\frac{t}{i\hbar}\delta_{\mathcal{H}}\right)\right)\int_{t_0}^{t}dt'\left(\exp\left(-\frac{t'}{i\hbar}\delta_{\mathcal{H}}\right)\right)\frac{1}{i\hbar}\delta_{\mathcal{H}}\rho_0\notag
\\
&=-\left(\exp\left(\frac{t}{i\hbar}\delta_{\mathcal{H}}\right)\right)\int_{t_0}^{t}dt'\left(\frac{d}{dt'}\exp\left(-\frac{t'}{i\hbar}\delta_{\mathcal{H}}\right)\right)\rho_0\notag
\\
&=\left(\exp\left(\frac{t-t_0}{i\hbar}\delta_{\mathcal{H}}\right)\right)\rho_0-\rho_0\notag
\\
&=\frac{1}{Z_0(\beta)}\exp\left( -\beta\mathcal{H}_0-\beta\mathcal{R}(t,t_0) \right)-\rho_0=\left(\Delta \rho(t)\right)_{\text{Z-type}},\label{eq:32}
\end{align}
where $\delta_{\mathcal{H}}$ denotes the inner derivation[19-21] defined by $\delta_{\mathcal{H}}Q=[\mathcal{H},Q]=\mathcal{H}Q-Q\mathcal{H}$, and we have also used the relations $\delta_{\mathcal{H}_1}\rho_0=\delta_{\mathcal{H}}\rho_0$ and 
\begin{equation}
\left(\exp\left(\frac{t-t_0}{i\hbar}\delta_{\mathcal{H}}\right)\right)\mathcal{H}_0=\int_{t_0}^{t}dt'\left(\exp\left(\frac{t-t'}{i\hbar}\delta_{\mathcal{H}}\right)\right)\frac{1}{i\hbar}[\mathcal{H}_1,\mathcal{H}_0]+\mathcal{H}_0=\mathcal{R}(t,t_0)+\mathcal{H}_0.\label{eq:33}
\end{equation} 
This transformation is easily extended to the case of the general time-dependent field $\Vec{F}(t)$.

\section{Mechanical formulation of thermal disturbance to derive entropy production}
In order to apply the general formulations based on the von Neumann equation with the Hamiltonian form (\ref{eq:6}) presented in the preceding sections and to perform the first-principles derivation of entropy production even for thermal conduction, we have to formulate thermal disturbance in a mechanical way.

The present theory \cite{1} about entropy production is based on the von Neumann equation (\ref{eq:6}) for density matrix $\rho(t)$ and for the Hamiltonian $\mathcal{H}(t)$.
The density matrix $\Delta \rho (t)$ in Eq.(\ref{eq:8}) is statistical through $\rho_0$, that is, it may contain the temperature $T$.
Now we extend our interpretation of the von Neumann equation (\ref{eq:8}) so that we may treat thermal conductivity using the mechanical perturbation method in Kubo's paper\cite{2}.
For this purpose, the Hamiltonian for thermal disturbance should be regarded to be an effective one renormalized in a local but enough large region to catch up the temperature gradient\cite{8,9}, as in the Ginzburg-Landau-Wilson Hamiltonian describing critical phenomena.
Then, we introduce a thermal field $\Vec{E}_{\text{T}}=\nabla \beta(\Vec{r})/\beta_0$ as a field $\Vec{F}(t)$ in Eq.(\ref{eq:6}), where $\beta(\Vec{r})=1/k_{\text{B}}T(\Vec{r})$ and $\beta_0$ denotes the characteristic $\beta$.
Such a renormalized particle as moves receiving a force from the field $\Vec{E}_{\text{T}}$ is called "a thermal particle".
The next problem is to find an operator $\Vec{A}_{\text{H}}$ conjugate to the thermal field $\Vec{E}_{\text{T}}(t)$.
The operator $\Vec{A}_{\text{H}}$ should be chosen so that the time-derivative $\Vec{\dot{A}}_{\text{H}}$ may give the heat current $\Vec{j}_{\text{H}}$, namely $\Vec{\dot{A}}_{\text{H}}=\Vec{j}_{\text{H}}$.
The heat current $\Vec{j}_{\text{H}}$ is first defined by
\begin{equation}
\Vec{j}_{\text{H}}=\sum_{j}\{ h_j \Vec{v}_j \}=\frac{1}{2}\sum_{j}(h_j \Vec{v}_j+ \Vec{v}_j h_j)\label{eq:34}
\end{equation}
in the simplest case of ideal renormalized particles, when the unperturbed Hamiltonian $\mathcal{H}_0$ is given in the form
\begin{equation}
\mathcal{H}_0=\sum_{j} h_j;\quad h_j=\epsilon_j +u_j \quad \text{ with } \quad \epsilon_j=\frac{1}{2m^{*}}\Vec{p}^2_j \quad \text{ and }\quad u_j=u(\Vec{r}_j).\label{eq:35}
\end{equation}
Here the last quantity $u_j$ denotes the scattering by impurities, and $m^{*}$ denotes the effective mass defined by Eq.(\ref{eq:39}).
It should be also noted that $[h_j,h_k]=0$ or $[\epsilon_k,u_j]=0$ for $j\not=k$.
The time-derivative $\Vec{\dot{A}}_{\text{H}}$ is defined by
\begin{equation}
\Vec{\dot{A}}_{\text{H}}=\frac{1}{i\hbar}[\Vec{A}_{\text{H}},\mathcal{H}_0],\label{eq:36}
\end{equation}
as usual.
It is easily proved that the heat operator $\Vec{A}_{\text{H}}$ of the form
\begin{equation}
\Vec{A}_{\text{H}}=\sum_{j}\{ h_j \Vec{r}_j \}\equiv \frac{1}{2}\sum_{j}(h_j \Vec{r}_j+ \Vec{r}_j h_j)\label{eq:37}
\end{equation}
satisfies the required relation
\begin{equation}
\Vec{\dot{A}}_{\text{H}}\equiv\frac{1}{i\hbar}[\Vec{A}_{\text{H}},\mathcal{H}_0]=\sum_{j}\{ h_j \Vec{v}_j\}\equiv\Vec{j}_{\text{H}}.\label{eq:38}
\end{equation}
Here we have used the renormalized momentum $\Vec{p}_j$ and space coordinate $\Vec{r}_j$ of the effective particle $j$ defined in a local region of original $n$ particles as follows:
\begin{equation}
\Vec{p}_j=\sum_{i=1}^{n}\Vec{p}_{ji},\quad\Vec{r}_j=\frac{1}{n}\sum_{i=1}^{n}\Vec{r}_{ji},\quad m^{*}=nm.\label{eq:39}
\end{equation}
We have also used the following commutation relations
\begin{equation}
[\Vec{r}_j,\Vec{p}_j]=\frac{1}{n}\sum_{i=1}^{n}[\Vec{r}_{ji},\Vec{p}_{ji}]=i\hbar\quad\text{ and }\quad\Vec{\dot{r}}_j=\Vec{v}_j,\label{eq:40}
\end{equation}
which have been derived from Eq.(\ref{eq:39}).

Thus, we have found a heat operator $\Vec{A}_{\text{H}}$ conjugate to the thermal field $\Vec{E}_{\text{T}}$ in the form (\ref{eq:37}).
Once it is found, we may use it in a generalized case in which interactions among particles exist, that is, $u_j$ is given by $u_j=u(\Vec{r}_j,\{\Vec{r}_k\})$ and consequently we have $[h_j,h_k]\not=0$ or $[\epsilon_j,u_j]\not=0$ for such $\{k\}$ as the $j$-th and $k$-th particles interact.
In this general case, the heat current is easily shown to be given in the form
\begin{equation}
\Vec{j}_{\text{H}}=\sum_{j}\{ h_{j} \Vec{v}_{j}\}+\frac{1}{2i\hbar}\sum_{i,j}\{(\Vec{r}_j-\Vec{r}_k)[u_j(\Vec{r}_j,\{\Vec{r}_k\}),\epsilon_k-\epsilon_j] \}.\label{eq:41}
\end{equation}

Now we expand $\rho(t)$ in the von Neumann equation (\ref{eq:6}) as
\begin{equation}
\rho_{\text{lr}}(t)=\rho_0+\rho_1(t);\quad\rho_0=e^{-\beta\mathcal{H}_0}/Z_0(\beta)\label{eq:42}
\end{equation}
with $Z_0(\beta)={\text{Tr}}\exp(-\beta\mathcal{H}_0)$, up to the first order of the thermal field $\Vec{E}_{\text{T}}$.
Then, the first-order density matrix $\rho_1(t)$ is expressed as
\begin{equation}
\rho_1(t)=\rho_0\int_{t_0}^{t}ds\int_{0}^{\beta}d\lambda \Vec{E}_{\text{T}}\cdot\Vec{\dot{A}}_{\text{H}}(s-t-i\hbar\lambda), \label{eq:43}
\end{equation}
where $\Vec{\dot{A}}_{\text{H}}$ denotes the current $\Vec{j}_{\text{H}}$ of $\Vec{A}_{\text{H}}$, namely $\Vec{j}_{\text{H}}=\Vec{\dot{A}}_{\text{H}}=[\Vec{A}_{\text{H}},\mathcal{H}_0]/i\hbar$.
Thus, we obtain the well-known formula\cite{2,9}
\begin{align}
\Vec{J}_{\text{H}}\equiv \langle \Vec{j}_{\text{H}}\rangle _t &= {\text{Tr}}\left\{ \left( \rho_0+\rho_1(t) \right) \Vec{j}_{\text{H}} \right\}\equiv \kappa\Vec{E}_{\text{T}}, \notag
\\
\kappa&= \int_{0}^{\infty}ds\int_{0}^{\beta}d\lambda e^{-\epsilon s} \langle  \Vec{j}_{\text{H}}\Vec{j}_{\text{H}}(s+i\hbar \lambda) \rangle _0 \Vec{E}_{\text{T}}\label{eq:44}
\end{align}
by taking the limit $t_0\rightarrow -\infty$ for the initial $t_0$ after the thermodynamic limit.

The present formulation of thermal conductivity is similar to Luttinger's theory\cite{10} based on a gravitational field which causes energy or heat current to flow.
(In his theory, an energy density $h(\Vec{r})$ is regarded to behave as if it had a mass density $h(\Vec{r})/c^2$ for light velocity $c$ and a very weak gravitational field is induced\cite{11,12} by the inhomogeneity of temperature.
Of course, Luttinger used this idea only conceptually in his formulation.)
However, our formulation of introducing the above thermal field $\Vec{E}_{\text{T}}$ and its conjugate heat operator $\Vec{A}_{\text{H}}$ through the idea of renormalization of semi-local thermal effects is more transparent for deriving the transport coefficient (\ref{eq:44}) of the Kubo form.
Thus we have arrived at a unified "mechanical" formulation of linear responses including thermal conductivity\cite{2,9}.

In particular, to take into account the supply of heat energy is crucial to keep the temperature gradient in the thermal conduction.
This mechanism of energy supply will be formulated in Sections 7 and 8.

\section{Entropy production of thermal disturbance}
We discuss here the entropy production of thermal disturbance using the above Kubo formula (\ref{eq:44}).
As for the static electric conduction, the entropy production for the thermal disturbance is given similarly in the form
\begin{equation}
\left(\frac{dS}{dt}\right)_{\text{irr}}=\frac{1}{T}\text{Tr}\mathcal{H}_0\rho'_2(t)=\frac{\Vec{J}_{\text{H}}\cdot \Vec{E}_{\text{T}}}{T}=\frac{\kappa E^2_{\text{T}}}{T}\label{eq:45}
\end{equation}
per unit volume and per unit time, as far as the lowest order of $E_{\text{T}}$ is concerned.
In general, we have 
\begin{equation}
\left(\frac{dS}{dt}\right)_{\text{irr}}(E_{\text{T}})=\frac{\kappa (E_{\text{T}}) E^2_{\text{T}}}{T(t)}\label{eq:46}
\end{equation}
for an arbitrary strength of the thermal field.
These formulas have been derived here on the basis of the above mechanical formulation.

In contrast to the electric conduction, the converse argument on the entropy production, namely to discuss it thermodynamically after the firs-principles derivation of it will be very instructive in order to understand physically the mechanism of the entropy production for the thermal disturbance.

The irreversible entropy production per unit volume and per unit time is also calculated by
\begin{equation}
\left( \frac{dS}{dt} \right)_{\text{irr}}=\frac{1}{L}J_{\text{H}}\left( \frac{1}{T}-\frac{1}{T+\Delta T} \right)=\frac{J_{\text{H}}E_{\text{T}}}{T}\label{eq:47}
\end{equation}
in the lowest order of $E_{\text{T}}$, where $E_{\text{T}}=(\Delta T/L)/T$, namely $\Delta T=LTE_{\text{T}}$ for the system size $L$.
The heat energy supplied from outside to keep the thermal field or thermal gradient (as has been discussed in Section 5) is transformed to the relevant system, that is, the energy$\left< \mathcal{H}_1 \right>_t=-\left< \Vec{A}_{\text{H}}\cdot\Vec{E}_{\text{T}} \right>_t$ is changed into the internal energy $\left< \mathcal{H}_0 \right>_t$, whose increase yields the entropy production
\begin{equation}
\left( \frac{dS}{dt} \right)_{\text{irr}}=\frac{1}{T}\frac{d\left< \mathcal{H}_0 \right>_t}{dt}=\frac{\kappa(E_{\text{T}})E^2_{\text{T}}}{T},\label{eq:48}
\end{equation}
as in the electric conduction.
This gives a unified explanation of the entropy production both for the electric conduction and thermal conduction.

The stationary state for the thermal disturbance can also be formulated\cite{1} using the symmetry-separated von Neumann equation, as in the case of the electric conduction.

\section{Symmetry-separated von Neumann equation with energy supply and new scheme to express stationary states}
In the previous paper\cite{1}, we have introduced the symmetry-separated von Neumann equation in order to take into account the effect of extracting the generated heat outside and to keep the internal energy $\left< \mathcal{H}_0 \right>_t$ constant.
In the present and succeeding sections, we assume again that the operator $\mathcal{H}_1$ is time-independent.
However, the energy $\left< \mathcal{H}_1 \right>_t $ decreases as time $t$ increases.
In this section, we try to find a new formulation in which both $\left< \mathcal{H}_0 \right>_t$ and $\left< \mathcal{H}_1 \right>_t$ are time-independent, while there exists the current $\Vec{J}=\text{Tr}\Vec{j}\rho(t)$ and the entropy production $(dS/dt)_{\text{irr}}$ is positive, even in steady states.
For this purpose, we extend here the concept of a steady state by defining the corresponding stationary density-matrix $\rho^{\text{(st)}}(t)$ as
\begin{equation}
\frac{d}{dt}\rho^{\text{(st)}}(t)={\text{constant operator}},\label{eq:49}
\end{equation} 
namely, $\rho^{\text{(st)}}(t)$ may contain a linear part with respect to time $t$ as
\begin{equation}
\rho^{\text{(st)}}(t)=\rho (0)+t\rho' (0)\label{eq:50}
\end{equation}
for the density matrix $\rho(0)$ defined in the limit $t_0\rightarrow -\infty$.
(For $t\rightarrow t_0 \rightarrow -\infty$, the above $t$ should be interpreted as $te^{\epsilon t}$ with $\epsilon\rightarrow +0$ and similarly in Eqs.(\ref{eq:74}) and (\ref{eq:75}).)

The main problem here is to find an extended von Neumann equation to describe the process of energy supply without changing the previous physical results\cite{1} on $\left< \mathcal{H}_0 \right>_t(=\text{constant})$ and $(dS/dt)_{\text{irr}}(>0)$ except $\left< \mathcal{H}_1 \right>_t$ (which decreases as time $t$ increases in the previous treatment\cite{1}).
Only one additional condition of the present new formulation is to require that $\left< \mathcal{H}_1 \right>_t$ is also constant.

Our strategy to accomplish this scenario is the following:
\begin{enumerate}
\item[(i)] First notice that the idea\cite{1} of symmetry separation of the density matrix $\rho(t)$ as $\rho(t)=\rho_{\text{s}}(t)+\rho_{\text{a}}(t)$ plays a crucial role in formulating energy supply from outside to the relevant system as well as in formulating the extraction of the generated heat\cite{1}.
In fact, the effect of energy supply is expressed in the partial von Neumann equation concerning the antisymmetric part $\rho_{\text{a}}(t)$ as 
\begin{equation}
\frac{\partial \rho_{\text{a}}(t)}{\partial t}=\frac{1}{i\hbar}[\mathcal{H}_0,\rho_{\text{a}}(t)]+\frac{1}{i\hbar}[\mathcal{H}_1,\rho_{\text{s}}(t)]-\epsilon\rho_{\text{a}}(t)+\eta_{\text{es}}(t),\label{eq:51}
\end{equation}
where $\eta_{\text{es}}(t)$ denotes the energy source term.

\item[(ii)] The energy source term $\eta_{\text{es}}(t)$ should be determined so that the solution $\rho_{\text{a}}(t)$ of Eq.(\ref{eq:51}) coupled with the following equation
\begin{equation}
\frac{\partial \rho_{\text{s}}(t)}{\partial t}=\frac{1}{i\hbar}[\mathcal{H}_0,\rho_{\text{s}}(t)]+\frac{1}{i\hbar}[\mathcal{H}_1,\rho_{\text{a}}(t)]-\epsilon_{\text{r}}(\rho_{\text{s}}(t)-\rho_{0})\label{eq:52}
\end{equation}
may satisfy the conditions:
\begin{enumerate}
\item[(a)] The contribution of the source term $\eta_{\text{es}}(t)$ to the derivative $d\left< \mathcal{H}_1 \right>_t/dt$, namely $\text{Tr}\mathcal{H}_1\eta_{\text{es}}(t)$ should be non-vanishing.

\item[(b)] The other quantities $\left< \mathcal{H}_0 \right>_t=\text{Tr}\mathcal{H}_0\rho_{\text{s}}(t)$, $\Vec{J}_F=\text{Tr}\Vec{j}\rho_{\text{a}}(t)$ and $(dS/dt)_{\text{irr}}$ should be the same as in the case of no energy supply (namely, $\eta_{\text{es}}(t)\equiv0$).
\end{enumerate}
Can we find such a desirable operator $\eta_{\text{es}}(t)$?

\item[(iii)] To answer the above question on $\eta_{\text{es}}(t)$, we try to study what kind of the density matrix $\tilde{\rho}(t)$ satisfies the above two conditions (i) and (ii).
\end{enumerate}

At first sight, this problem seems difficult.
However, we find the following favorable relations:
\begin{enumerate}
\item[(a)] For any operator function $\tilde{\rho}_0\equiv\tilde{\rho}_0(\mathcal{H}_0)$, we have
\begin{equation}
\text{Tr}\Vec{j}\{\mathcal{H}_1\tilde{\rho}_0\}=0;\quad \{ AB\}=\frac{1}{2}\{ AB+BA \},\label{eq:53}
\end{equation}
and more generally for any operator function $\tilde{\rho}_{\mu}\equiv\tilde{\rho}_{\mu}(\mathcal{H}_{0,\mu})$ with $\mathcal{H}_{o,\mu}\equiv\mathcal{H}_0+\mu\mathcal{H}_1$, we have
\begin{equation}
\text{Tr}\Vec{j}\{ \mathcal{H}^n_1 \tilde{\rho}_{\mu} \}=0;\quad \{ A^n B\}=\frac{1}{n+1}(A^n B+A^{n-1}BA+\cdots +BA^n)\label{eq:54}
\end{equation}
for any positive integer $n$ and for any real value of $\mu$, where $\Vec{j}=\Vec{\dot{A}}=[\Vec{A},\mathcal{H}_0]/(i\hbar)$ and $-\Vec{j}\cdot\Vec{F}=[\mathcal{H}_1,\mathcal{H}_0]/(i\hbar)$.
Equation (\ref{eq:53}) corresponds to the identity
\begin{equation}
\text{Tr}\left( [\mathcal{H}_1,\mathcal{H}_0]\{ \mathcal{H}_1 \tilde{\rho}_{\mu} \} \right) = \text{Tr}\left([\mathcal{H}_1,\mathcal{H}_{0,\mu}]\{ \mathcal{H}_1\tilde{\rho}_{\mu} \} \right)=0,  \label{eq:55}
\end{equation}
which can be easily proven using the rule $\text{Tr}ABC=\text{Tr}BCA$, and using $[\mathcal{H}_{0,\mu},\tilde{\rho}_\mu]=0$.
Equation (\ref{eq:54}) can also be proven similarly.
\end{enumerate}
Note also the following relations
\begin{enumerate}
\item[(b)] For $\mu=0$, we have
\begin{equation}
\text{Tr}(\mathcal{H}_0\{ \mathcal{H}_1 \tilde{\rho}_0 \})=0,\label{eq:56}
\end{equation}
and more generally
\begin{equation}
\text{Tr}(\mathcal{H}_0\{ \mathcal{H}^{2n-1}_1 \tilde{\rho}_0 \})=0,\label{eq:57}
\end{equation}
for any positive integer $n$, as is easily seen from symmetry.
\end{enumerate}

These relations suggest that the solution of Eq.(\ref{eq:51}) will take the possible form
\begin{equation}
\rho_{\text{a}}(t)=\hat{\rho}_{\text{a}}(t)+\tilde{\rho}_{\text{a}}(t);\quad \tilde{\rho}_{\text{a}}(t)=\epsilon_{\text{es}}(t)\{ \mathcal{H}_1 \tilde{\rho}_0 \},\label{eq:58}
\end{equation}
more generally
\begin{equation}
\tilde{\rho}_{\text{a}}(t)=\epsilon_{\text{es}}(t)\{ \mathcal{H}_1^{2n-1} \tilde{\rho}_0 \},\label{eq:59}
\end{equation}
where $\hat{\rho}_{\text{a}}(t)$ is the solution of Eq.(\ref{eq:51}) without the source term $\eta_{\text{es}}(t)$.
From the condition $\tilde{\rho}_{\text{a}}(t_0)=0$, we have $\epsilon_{\text{es}}(t_0)=0$.
Note also that the trace of the total density matrix
\begin{equation}
\rho(t)=\hat{\rho}(t)+\tilde{\rho}(t)\quad \text{ with }\quad \tilde{\rho}(t)=\tilde{\rho}_{\text{a}}(t)+\tilde{\rho}_{\text{s}}(t)\label{eq:60}
\end{equation}
is conserved, namely $\text{Tr}\rho(t)=1$, when $\tilde{\rho}_{\text{a}}(t)$ takes the form (\ref{eq:58}) for any function of $\epsilon_{\text{es}}(t)$.
Here, $\hat{\rho}_{\text{s}}(t)$ and $\hat{\rho}_{\text{a}}(t)$ are the solutions of the following coupled equations without the source term $\eta_{\text{es}}(t)$:
\begin{align}
\frac{\partial }{\partial t}\hat{\rho}_{\text{s}}(t)&=\frac{1}{i\hbar}[\mathcal{H}_0,\hat{\rho}_{\text{s}}(t)]+\frac{1}{i\hbar}[\mathcal{H}_1,\hat{\rho}_{\text{a}}(t)]-\epsilon_{\text{r}}(\hat{\rho}_{\text{s}}-\rho_0),\label{eq:61}
\\
\frac{\partial }{\partial t}\hat{\rho}_{\text{a}}(t)&=\frac{1}{i\hbar}[\mathcal{H}_0,\hat{\rho}_{\text{a}}(t)]+\frac{1}{i\hbar}[\mathcal{H}_1,\hat{\rho}_{\text{s}}(t)]-\epsilon\hat{\rho}_{\text{a}}(t).\label{eq:62}
\end{align}
The tilde density matrices $\tilde{\rho}_{\text{s}}$ and $\tilde{\rho}_{\text{a}}$ denote the contribution from the source term $\eta_{\text{es}}(t)$ and they satisfy the following coupled equations:
\begin{align}
\frac{\partial }{\partial t}\tilde{\rho}_{\text{s}}(t)&=\frac{1}{i\hbar}[\mathcal{H}_0,\tilde{\rho}_{\text{s}}(t)]+\frac{1}{i\hbar}[\mathcal{H}_1,\tilde{\rho}_{\text{a}}(t)],\label{eq:63}
\\
\frac{\partial }{\partial t}\tilde{\rho}_{\text{a}}(t)&=\frac{1}{i\hbar}[\mathcal{H}_0,\tilde{\rho}_{\text{a}}(t)]+\frac{1}{i\hbar}[\mathcal{H}_1,\tilde{\rho}_{\text{s}}(t)]+\eta_{\text{es}}(t).\label{eq:64}
\end{align}

Then, what is the source term $\eta_{\text{es}}(t)$ to give the solution of the form (\ref{eq:58}) (or (\ref{eq:59})) in the above coupled equations (\ref{eq:63}) and (\ref{eq:64})?
The answer is the following:
\begin{equation}
\eta_{\text{es}}(t)=\epsilon'_{\text{es}}(t)\{ \mathcal{H}_1 \tilde{\rho}_0 \} - \frac{1}{i\hbar}\epsilon_{\text{es}}(t)[\mathcal{H}_0,\{ \mathcal{H}_1 \tilde{\rho}_0 \}] - \frac{1}{i\hbar}[\mathcal{H}_1,\tilde{\rho}_{\text{s}}(t)],\label{eq:65}
\end{equation}
where
\begin{equation}
\tilde{\rho}_{\text{s}}(t)=\frac{1}{i\hbar}\int_{t_0}^{t}e^{\frac{(t-s)}{i\hbar}\mathcal{H}_0}[\mathcal{H}_1,\tilde{\rho}_{\text{a}}(s)]e^{-\frac{(t-s)}{i\hbar}\mathcal{H}_0}ds, \quad \text{and} \quad \tilde{\rho}_{\text{a}}(t)=\epsilon_{\text{es}}(t)\{ \mathcal{H}_1\tilde{\rho}_{0} \}.\label{eq:66}
\end{equation}
We have only solved inversely Eqs.(\ref{eq:63}) and (\ref{eq:64}).
The parameter $\epsilon_{\text{es}}(t)$ is taken to be
\begin{equation}
\epsilon_{\text{es}}(t)=(t+a)e^{\epsilon t}\epsilon'_{\text{es}}(0)\label{eq:67}
\end{equation}
for $\epsilon\rightarrow+0$ and for a constant $a$ satisfying the relation $a=\epsilon_{\text{es}}(0)/\epsilon'_{\text{es}}(0)$, in order to describe a steady state, as will be discussed in the next section.
The parameter (\ref{eq:67}) has the desired properties that $\epsilon_{\text{es}}(t)$ is essentially linear in time $t$ and $\epsilon_{\text{es}}(-\infty)=0$.

Thus, we have finally obtained our basic equations to describe our desired non-equilibrium phenomena with energy supply from outside to the system and with the extraction of generated heat outside.
One may say that an electric field $\Vec{E}$ in $\mathcal{H}_1$ for electric conduction should be regarded to be an external force and consequently that the above energy supply to $\mathcal{H}_1$ may be unnecessary.
However, it seems reasonable in the case of thermal conductance to consider heat energy supply.

\section{True stationary states with both energy supply and heat extraction}
First we express our basic equation obtained in Section 7 in the following compact form
\begin{equation}
\frac{d}{dt}\Vec{\rho}(t)=-\mathcal{L}\Vec{\rho}(t)+\mathcal{L}_{\text{s}}(t)\Vec{\rho}_0.\label{eq:68}
\end{equation}
Here, we have introduced the vector $\Vec{\rho}(t)$ and the super-matrices $\mathcal{L}$ and $\mathcal{L}_{\text{s}}(t)$:
\begin{equation}
\Vec{\rho}(t)=
\begin{pmatrix}
\rho_{\text{s}}(t) 
\\
\rho_{\text{a}}(t) 
\end{pmatrix},\quad
\mathcal{L}=
\begin{pmatrix}
\epsilon_{\text{r}}\mathcal{A} & -\epsilon_{\text{r}}\mathcal{B}
\\
-\epsilon \mathcal{D} & \epsilon \mathcal{C}
\end{pmatrix},\quad
\mathcal{L}_{\text{s}}(t)=
\begin{pmatrix}
\epsilon_{\text{r}} & 0
\\
0 & \mathcal{L}^{(22)}_{\text{s}}(t)
\end{pmatrix},
\label{eq:69}
\end{equation}
where the super-operators $\mathcal{A},\mathcal{B},\mathcal{C}$ and $\mathcal{D}$ are defined \cite{1} by
\begin{align}
&\mathcal{A}=1-\omega_{\text{r}}\delta_{\mathcal{H}_0},\quad \mathcal{B}=\omega_{\text{r}}\delta_{\mathcal{H}_1},\quad \mathcal{C}=1-\omega_{\epsilon}\delta_{\mathcal{H}_0},\quad \mathcal{D}=\omega_{\epsilon}\delta_{\mathcal{H}_1},\notag
\\
&\omega_{\text{r}}=\frac{1}{i\hbar\epsilon_{\text{r}}},\quad \omega_{\epsilon}=\frac{1}{i\hbar\epsilon},\quad \delta_{A}Q\equiv[A,Q]=AQ-QA.\label{eq:70}
\end{align}
Furthermore, the vector $\Vec{\rho}_0$ and the super-operator $\mathcal{L}^{(22)}_{\text{s}}(t)$ are defined as
\begin{align}
\Vec{\rho}_0=
\begin{pmatrix}
\rho_0
\\
\tilde{\rho}_0
\end{pmatrix}
\quad \text{and}\quad
\mathcal{L}^{(22)}_{\text{s}}(t)\tilde{\rho}_0 =\eta_{\text{es}}(t).\label{eq:71}
\end{align}
(Here we may have put $\tilde{\rho}_0=\rho_0$ for simplicity.
The unique condition on $\tilde{\rho}_0$ is the commutativity of it with $\mathcal{H}_0$, namely $[\mathcal{H}_0,\tilde{\rho}_0]=0$.)
Then, the energy source operator $\eta_{\text{es}}(t)$ given by Eq.(\ref{eq:65}) with Eq.(\ref{eq:66}) is linear in $\rho_0$, and consequently the super operator $\mathcal{L}^{(22)}_{\text{s}}(t)$ can be defined as above.
The formal solution of Eq.(\ref{eq:68}) is given by
\begin{equation}
\Vec{\rho}(t)=\int_{t_0}^{t}e^{-(t-t')\mathcal{L}}\mathcal{L}_{\text{s}}(t')\Vec{\rho}_0dt'+e^{-(t-t_0)\mathcal{L}}
\begin{pmatrix}
\rho_0
\\
0
\end{pmatrix}.\label{eq:72}
\end{equation}
It should be remarked that the above $\mathcal{L}$ and $\mathcal{L}_{\text{s}}(t)$ are super matrices whose components are super operators such as $\mathcal{A},\mathcal{B},\mathcal{C}$ and $\mathcal{D}$.

In the limit $t_0\to -\infty$, Eq.(\ref{eq:72}) can be transformed into the following "stationary" solution
\begin{equation}
\Vec{\rho}^{\text{(st)}}(t)=\int_{0}^{\infty}e^{-s\mathcal{L}}\mathcal{L}_{\text{s}}(t-s)\Vec{\rho}_0ds.\label{eq:73}
\end{equation}
This can be expressed in the form
\begin{equation}
\Vec{\rho}^{\text{(st)}}(t)=\Vec{\rho}(0)+t\Vec{\rho}'(0)\label{eq:74}
\end{equation}
together with the formula
\begin{equation}
\mathcal{L}_{\text{s}}(t)\Vec{\rho}_0=\mathcal{L}_{\text{s}}(0)\Vec{\rho}_0+t\mathcal{L}'_{\text{s}}(0)\Vec{\rho}_0,\label{eq:75}
\end{equation}
which are correct for any $t$ as is seen from Eq.(\ref{eq:67}) and Eq.(\ref{eq:71}) with Eq.(\ref{eq:65}) for $t_0\rightarrow -\infty$ and $\epsilon\rightarrow+0$.
Here we have
\begin{align}
\Vec{\rho}(0)&=\int_0^{\infty}e^{-t\mathcal{L}}\mathcal{L}_{\text{s}}(-t)\Vec{\rho}_0dt\notag
\\
&=\int_0^{\infty}e^{-t\mathcal{L}}\mathcal{L}_{\text{s}}(0)\Vec{\rho}_0dt-\int_0^{\infty}e^{-t\mathcal{L}}\mathcal{L}'_{\text{s}}(0)\Vec{\rho}_0 tdt\notag
\\
&=\mathcal{L}^{-1}\left( \mathcal{L}_{\text{s}}(0)\Vec{\rho}_0-\mathcal{L}^{-1}\mathcal{L}'_{\text{s}}(0)\Vec{\rho}_0 \right),\label{eq:76}
\end{align}
and
\begin{equation}
\Vec{\rho}'(0)=\int_{0}^{\infty}e^{-t\mathcal{L}}\mathcal{L}'_{\text{s}}(0)\Vec{\rho}_0dt =\mathcal{L}^{-1}\mathcal{L}'_{\text{s}}(0)\Vec{\rho}_0.\label{eq:77}
\end{equation}
Note that $\text{Tr}\Vec{\rho}'(0)=0$ and consequently that $\text{Tr}\rho(t)=\text{Tr}\rho_{\text{s}}(t)=1$, as it should be.
It should be noted that our extended definition of the stationary density matrix (\ref{eq:74}) does not violate the ordinary stationarity of the current and $\rho_{\text{s}}(t)$ except the energy supply proportional to time. 
(See also Appendix for the inverse $\mathcal{L}^{-1}$ of the super-matrix $\mathcal{L}$ whose elements are super-operators.)
Equivalently we have
\begin{equation}
\mathcal{L}\Vec{\rho}(0)=\mathcal{L}_{\text{s}}(0)\Vec{\rho}_{0}-\Vec{\rho}'(0),\quad\text{and}\quad \mathcal{L}\Vec{\rho}'(0)=\mathcal{L}'_{\text{s}}(0)\Vec{\rho}_0.\label{eq:78}
\end{equation}
These relations can also be obtained directly from Eq.(\ref{eq:68}), by expressing $\mathcal{L}_{\text{s}}(t)$ as in Eq.(\ref{eq:75}).

Thus we have finally arrived at the formal solution of the steady state, namely the "stationary" density-matrix $\Vec{\rho}^{\text{(st)}}=\Vec{\rho}(0)+t\Vec{\rho}'(0)$ with Eqs.(\ref{eq:76}) and (\ref{eq:77}), according to our new definition (\ref{eq:49}).

The time-derivative of the energy $\left<\mathcal{H}_1\right>_1$ is also given by
\begin{align}
\frac{d}{dt}\left<\mathcal{H}_1\right>_t &=\text{Tr}\left(\mathcal{H}_1\eta_{\text{es}}(t)\right)-\Vec{J}_F\cdot\Vec{F}\notag
\\
&=\epsilon'_{\text{es}}(t)\text{Tr}\left(\mathcal{H}_1\{\mathcal{H}_1\tilde{\rho}_0 \}\right)-\Vec{J}_F\cdot\Vec{F}\notag
\\
&=\epsilon'_{\text{es}}(0)\text{Tr}\left( \mathcal{H}_1^2\tilde{\rho}_0 \right)-\Vec{J}_F\cdot\Vec{F},\label{eq:79}
\end{align}
using Eq.(\ref{eq:68}) together with Eqs.(\ref{eq:65}),(\ref{eq:67}),(\ref{eq:69}) and (\ref{eq:71}), and also using the relation (\ref{eq:53}).
Here, the nonlinear current $\Vec{J}_{F}$ is given by Eq.(\ref{eq:13}).
Therefore, by imposing the condition
\begin{equation}
\epsilon'_{\text{es}}(0)=\frac{\Vec{J}_F\cdot\Vec{F}}{\text{Tr}\left( \mathcal{H}_1^2 \tilde{\rho}_0\right)},\label{eq:80}
\end{equation}
we arrive at the desired conservation relation of $\left< \mathcal{H}_1 \right>_t$:
\begin{equation}
\left< \mathcal{H}_1 \right>_t=\text{constant}\label{eq:81}
\end{equation}
as well as the conservation of $\left< \mathcal{H}_0 \right>_t$, while the entropy production $\left(dS/dt\right)_{\text{irr}}=\Vec{J}_F\cdot\Vec{F}/T$ is positive.
This yields the true steady state with the irreversible (positive) entropy production.

In more general, the time derivative of the average $\left< Q \right>_t$ of any physical operator $Q=Q_{\text{s}}+Q_{\text{a}}$ for the symmetric part $Q_{\text{s}}$ and antisymmetric part $Q_{\text{a}}$ can be calculated by the formula:
\begin{align}
\frac{d}{dt}\left< Q \right>_t &=\frac{d}{dt}\left(\left< Q_{\text{s}} \right>_t+\left< Q_{\text{a}} \right>_t\right)\notag
\\
&=\frac{d}{dt}\left( \text{Tr}Q_{\text{s}}\rho_{\text{s}}(t)+\text{Tr}Q_{\text{a}}\rho_{\text{a}}(t) \right)\notag
\\
&=\frac{d}{dt}\text{Tr}\left( \Vec{Q}\cdot\Vec{\rho}(t) \right)\notag
\\
&=-\text{Tr}\left( \Vec{Q}\cdot\mathcal{L}\Vec{\rho}(t) \right)+\text{Tr}\left( \Vec{Q}\cdot\mathcal{L}_{\text{s}}\Vec{\rho}_0 \right),\label{eq:82}
\end{align}
using the basic equation (\ref{eq:68}) and the vector notation $\Vec{Q}$ defined by $\Vec{Q}^{\dagger}=(Q_{\text{s}},Q_{\text{a}})$.
For example, we have $\Vec{\mathcal{H}}^{\dagger}=(\mathcal{H}_0,\mathcal{H}_1)$ and $\Vec{\rho}^{\dagger}(t)=(\rho_{\text{s}}(t),\rho_{\text{a}}(t))$ with the inner product $(\Vec{\mathcal{H}}\cdot \Vec{\rho}(t))=\mathcal{H}_{\text{s}}\rho_{\text{s}}(t)+\mathcal{H}_{\text{a}}\rho_{\text{a}}(t)$.
The above formula (\ref{eq:82}) is convenient for studying the average of dynamical derivatives\cite{1} such as the entropy production in steady states, as was shown in Paper I.

\section{Summary and discussion}
The two types of nonlinear responses which have been formulated here look very different from one another, and in fact they give very different effects, if we make some approximations in the general expressions (\ref{eq:16}) and (\ref{eq:26}).
These two formulations yield the same physical quantities in different forms and consequently we obtain many identities on correlations \cite{25,26}.
These may be related to fluctuation theorems[27-31], which will be discussed somewhere else.
In particular, the present Zubarev-type formulation of nonlinear relaxation will be useful to discuss the relaxation of quantum systems to equilibrium[32-36].

When we treat the formal solution (\ref{eq:18}) with Eq.(\ref{eq:20}), we have to be careful about the boundedness of $\mathcal{R}(t,t_0)$.
It is often unbounded below.
Even the linear approximation of it, that is, Zubarev's expression (\ref{eq:31}) has some mathematical difficulty.
The operator $\mathcal{R}_1$ defined by Eq.(\ref{eq:29}) is unbounded, because the current operator $\Vec{j}$ is proportional to the momentum operator $\Vec{p}=\sum_j \Vec{p}_j$ in the electric conduction, and because the $\Vec{j}$ is proportional to $\{ \Vec{p}_j^3 \}$ in the thermal conduction, as is seen from Eq.(\ref{eq:41}).
However, $\mathcal{H}_0+\mathcal{R}_1$ is bounded below in the electric conduction, while it is still unfounded below in the thermal conduction.
Thus, the Kubo formula works but Zubarev's formula does not in the form (\ref{eq:30}) on the thermal conduction.
In any case, the perturbational  formulas of the transport coefficients up to some finite orders of the external force can be used safely, while some compact forms including  infinite order terms have to be treated carefully in this respect.
\begin{figure}[tbhp]
\begin{center}
\includegraphics[width=12cm,clip]{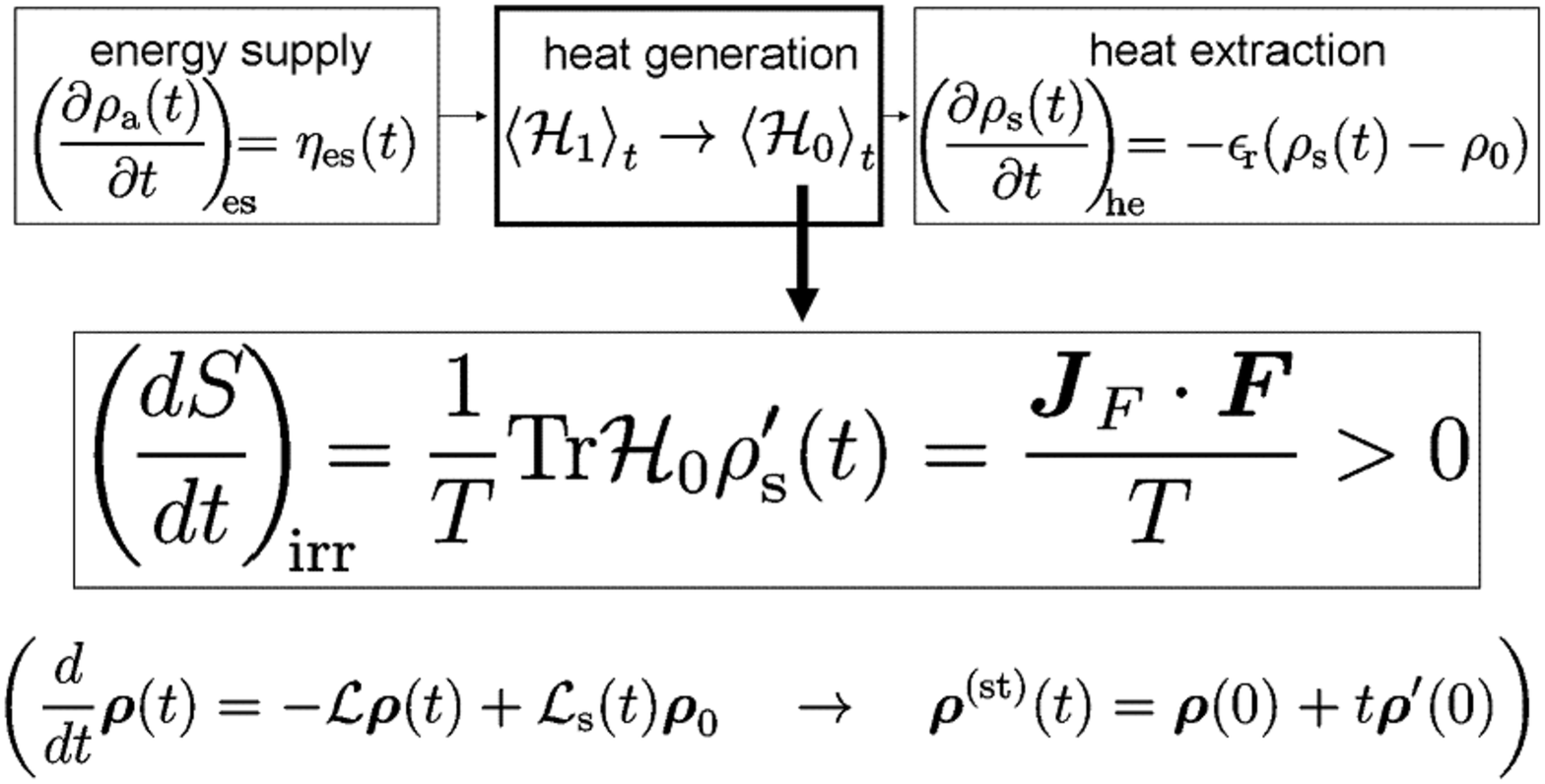}
\end{center}
\caption{Scheme of irreversible stationary transport phenomena (such as the electric conduction) with energy supply and heat extraction.Here, $\Vec{J}_F$ denotes the current in a field $\Vec{F}$, and $\rho_{\text{s}}(t)$ and $\rho_{\text{a}}(t)$ denote the symmetric and antisymmetric parts of the density matrix $\rho(t)$, respectively. The vector-operators $\Vec{\rho}(t)$ and $\mathcal{L}_\text{s}(t)\Vec{\rho}_0$ are defined by $\Vec{\rho}^{\dagger}(t)=(\rho_{\text{s}}(t), \rho_{\text{a}}(t))$ and $(\mathcal{L}_{\text{s}}(t)\Vec{\rho}_{0})^{\dagger}=(\epsilon_{\text{r}}\rho_0, \eta_{\text{es}}(t))$, respectively.}
\label{fig:1}
\end{figure}

We have formulated the scheme showing the mechanism of energy supply to the system from outside like a battery in electric conduction.
Thus, we have succeeded in constructing a "true steady state" with the constant $\left< \mathcal{H}_0 \right>_t$ and $\left< \mathcal{H}_1 \right>_t$ and with positive entropy production, as is shown in Fig.\ref{fig:1}.
The symmetrized form $\{ \mathcal{H}_1\tilde{\rho}_0\}$ in Eq.(\ref{eq:58}) is vital to express energy supply in contrast to the commutator $[\mathcal{H},\rho(t)]$ in the extended von Neumann equation\cite{37}. 
The present formulation of energy supply will be extended by using $\tilde{\rho}_{\mu}$ which is an arbitrary function of the operator $\mathcal{H}_{0,\mu}=\mathcal{H}_0+\mu\mathcal{H}_1$ introduced above Eq.(\ref{eq:54}).
In this extended case, the operator $\tilde{\rho}_{\mu}$ contains both symmetric and antisymmetric parts and so does the energy source term $\eta_{\text{es}}(t)$.
Therefore, the situation will become complicated.
For example, we have to separate $\eta_{\text{es}}(t)$ as $\eta_{\text{es}}(t)=\eta_{\text{s}}^{\text{(es)}}(t)+\eta_{\text{a}}^{\text{(es)}}(t)$ where $\eta_{\text{s}}^{\text{(es)}}(t)$ (and $\eta_{\text{a}}^{\text{(es)}}(t)$) denotes a symmetric (and an antisymmetric) part of $\eta_{\text{es}}(t)$.
Correspondingly, the time-evolution equation of $\tilde{\rho}_{\text{s}}(t)$ contains the energy source term $\eta_{\text{s}}^{\text{(es)}}(t)$.
Then, the energy $\left< \mathcal{H}_0 \right>_t$ changes owing to the effect of $\eta_{\text{s}}^{\text{(es)}}(t)$.
This additional energy should be extracted outside by increasing the parameter $\epsilon_{\text{r}}$ in Eq.(\ref{eq:52}), in order to retain the stationary state as before.
Furthermore, we have to be careful about the normalization $\text{Tr}\rho(t)=\text{Tr}\rho_{\text{s}}(t)=1$ in this extended case.
Namely we have to modify $\eta_{\text{s}}^{\text{(es)}}(t)$ such that $\text{Tr}\tilde{\rho}_{\text{s}}(t)=0$.
This extended formulation describes irreversible transport phenomena sustaining also the energy flow without dissipation, through the change of the antisymmetric density-matrix $\rho_{\text{a}}(t)$.
This will give a microscopic or statistical-mechanical explanation of "non-equilibrium thermodynamics"[38-42], which will be discussed elsewhere. 

It should also be remarked that the unboundedness of $\mathcal{H}_1$ may be relevant to the stationary condition (\ref{eq:80}).
In the case in which the quantity $\text{Tr}\left(\mathcal{H}_1^2\tilde{\rho}_0\right)$ diverges as the system size becomes infinite, we have to make $\epsilon_{\text{es}}$ infinitesimally small in order to satisfy the condition (\ref{eq:80}).
This remark on the unboundedness of $\mathcal{H}_1$ (and consequently of $\mathcal{H}=\mathcal{H}_0+\mathcal{H}_1$) is also related to the reason why the non-equilibrium operator $\exp(-\beta(\mathcal{H}_0+\mathcal{R}_1))$ with $\mathcal{R}_1$ defined by Eq.(\ref{eq:29}) works in Zubarev's theory instead of $\exp(-\beta(\mathcal{H}_0+\mathcal{H}_1))$ in order to describe transport phenomena.
This situation will become more clear, if we consider a magnetic system composed of localized spins $\{ \Vec{S}_{j} \}$.
The Hamiltonian $\mathcal{H}^{\text{(m)}}$ of this magnetic system given by $\mathcal{H}^{\text{(m)}}=\mathcal{H}_0+\mathcal{H}_1$ with $\mathcal{H}_1=-\Vec{A}^{\text{(m)}}\cdot\Vec{H}$ and $\Vec{A}^{\text{(m)}}=\mu_{\text{B}}\sum_j S_j^z$ is very different from the Hamiltonian of electric conduction $\mathcal{H}^{\text{(e)}}=\mathcal{H}_0+\mathcal{H}_1$ with $\mathcal{H}_1=-\Vec{A}^{\text{(e)}}\cdot\Vec{E}$ and $\Vec{A}^{\text{(e)}}=e\sum_j\Vec{r}_j$, in the following three respects:

(i) $\mathcal{H}^{\text{(m)}}$ describes an equilibrium state, while $\mathcal{H}^{\text{(e)}}$ describes a non-equilibrium state,
(ii) $\mathcal{H}^{\text{(m)}}$ is bounded, while $\mathcal{H}^{\text{(e)}}$ is unbounded,
and
(iii) $\Vec{\dot{A}}^{\text{(m)}}$ denotes the change of spins, namely spin relaxation\cite{43} (not spin transport), while $\Vec{\dot{A}}^{\text{(e)}}$ denotes the electric current, as has been discussed in Paper I.

One of the remarkable features of the present formulation on stationary states is not to use explicit forms of interactions between the system and outer systems (outside of Hilbert space[44-46] or two coupled systems\cite{47,48}), but to make use of only $\mathcal{H}_0$ and $\mathcal{H}_1$ with the form $\mathcal{H}_1=-\Vec{A}\cdot\Vec{F}$ giving a current $\Vec{j}=\Vec{\dot{A}}$, as in Kubo's theory on the general frame-work of linear transport phenomena.
This form of $\mathcal{H}_1$ has been shown in the present paper to be valid even for thermal conductance.

It is also very interesting to remark that the time derivative of the entropy production $\sigma_{S}(t)\equiv (dS/dt)_{\text{irr}}$ is derived from Eq.(\ref{eq:14}) as
\begin{equation}
\frac{d}{dt}\sigma_{S}(t)=\frac{1}{T}\int_{0}^{\beta}\!\!d\lambda \left<\Vec{j}(-i\hbar\lambda )\Vec{j}(t;\Vec{F})\right>_0 F^2,\label{eq:83}
\end{equation}
which explains the generalized principle of entropy production in the steady state \cite{49}.

First-principles derivations of dissipation and entropy production for general frequency-dependent external forces such as $\Vec{E}\cos(\omega t)$ and $\Vec{H}\cos(\omega t)$ will be reported in the near future\cite{49}.

\section*{Acknowledgment}
The author would like to thank Dr.Y.Hashizume for discussion and digitization of the present manuscript.

\appendix
\section{Inverse super-matrix $\mathcal{L}^{-1}$ of the super-matrix $\mathcal{L}$ whose elements are super-operators}
We explain here how to define the inverse super-matrix $\mathcal{L}^{-1}$ of the super-matrix of the form
\begin{equation}
\mathcal{L}=
\begin{pmatrix}
A&B
\\
D&C
\end{pmatrix}
,\label{eq:a1}
\end{equation}
where the positions of $C$ and $D$ are different from the ordinary case.
Here we assume that $A^{-1}$ and $C^{-1}$ exist but that $B^{-1}$ and $D^{-1}$ do not necessarily exist.
They correspond to $\mathcal{A},\mathcal{B},\mathcal{C}$ and $\mathcal{D}$ in Eq.(\ref{eq:70}), namely
\begin{equation}
A=\epsilon_{\text{r}}\mathcal{A},\quad B=-\epsilon_{\text{r}}\mathcal{B},\quad C=\epsilon\mathcal{C},\quad D=-\epsilon \mathcal{D}.\label{eq:a2}
\end{equation}
First we define $\mathcal{L}^{-1}$ by the relation $\mathcal{L}\mathcal{L}^{-1}=1$ (unit super-matrix), namely we have
\begin{equation}
\sum_j \mathcal{L}_{ij}(\mathcal{L}^{-1})_{jk}=\delta_{ik}\quad (i,j,k=1,2).\label{eq:a3}
\end{equation}
Then, by solving the two-dimensional equation (\ref{eq:a3}), the inverse super-matrix $\mathcal{L}^{-1}$ is easily obtained as
\begin{equation}
\mathcal{L}^{-1}=\left((\mathcal{L}^{-1})_{ij}\right)=
\begin{pmatrix}
\mathcal{K}A^{-1}&-\mathcal{K}A^{-1}BC^{-1}
\\
-C^{-1}D\mathcal{K}A^{-1}&C^{-1}D\mathcal{K}A^{-1}BC^{-1}+C^{-1}
\end{pmatrix}
,\label{eq:a4}
\end{equation}
where $\mathcal{K}$ is defined by $\mathcal{K}=(1-A^{-1}BC^{-1}D)^{-1}$.
The operator $\mathcal{K}$ is rewritten as $\mathcal{K}=(1-\mathcal{A}^{-1}\mathcal{B}\mathcal{C}^{-1}\mathcal{D})^{-1}$ in the correspondence relation (\ref{eq:a2}).
Finally, it should be remarked that the relation $\mathcal{L}^{-1}\mathcal{L}=1$, namely $\mathcal{L}^{-1}\mathcal{L}=\mathcal{L}\mathcal{L}^{-1}$ is easily confirmed by explicit calculations with use of be above expressions (\ref{eq:a4}) or by a mathematical argument.
(In fact, any solutions $\mathcal{X}_1$ and $\mathcal{X}_2$ of $\mathcal{L}\mathcal{X}_1=1$ and $\mathcal{X}_2\mathcal{L}=1$ satisfy the relation $\mathcal{X}_1=1\times \mathcal{X}_1=(\mathcal{X}_2\mathcal{L})\mathcal{X}_1=\mathcal{X}_2(\mathcal{L}\mathcal{X}_1)=\mathcal{X}_2\times 1=\mathcal{X}_2$ even for the super-matrices $\mathcal{L},\mathcal{X}_1$ and $\mathcal{X}_2$ in any dimensions.)

\bibliographystyle{model1a-num-names}
\bibliography{<your-bib-database>}






% \end{thebibliography}

\end{document}